\documentclass{aip-cp}

\usepackage[numbers]{natbib}
\usepackage{rotating}
\usepackage{graphicx}

\begin{document}

\title{The growth with energy of exclusive $J/\Psi$ and $\Upsilon$ photo-production cross-sections and BFKL evolution}

\author[aff1,aff2]{Martin Hentschinski}
\eaddress{hentschinski@correo.nucleares.unam.mx}

\affil[aff1]{
Facultad de Ciencias F\'isico Matem\'aticas,
Benem\'erita Universidad Aut\'onoma de Puebla,
Puebla 1152, Mexico}
\affil[aff2]{Instituto de Ciencias Nucleares,
Universidad Nacional Aut\'onoma de M\'exico, \\ 
Apartado Postal 70-543,
CDMX 04510, Mexico}

\maketitle

\begin{abstract}
  We investigate whether NLO BFKL evolution is capable to describe the
  energy dependence of the exclusive photo-production cross-section of
  vector mesons $J/\Psi$ and $\Upsilon$ on protons. Our description is based
  on available NLO BFKL fits of the proton impact factor in inclusive
  DIS, which allow us to construct the necessary scattering amplitude
  at zero momentum transfer $t=0$. Assuming an exponential drop-off with
  $t$, this result allows us to calculate the exclusive photoproduction
  cross-section. Comparing our results with both HERA data (measured
  by H1 and ZEUS collaborations in $ep$ collision) and LHC data
  (measured by ALICE, CMS and LHCb collaborations in ultra-peripheral
  $pp$ and $pPb$ collision) we find that our framework provides a very
  good description of the energy dependence of the $J/\Psi$ and $\Upsilon$
  photoproduction cross-section, providing therefore further evidence
  for BFKL evolution at the LHC. The available fits of the proton
  impact factor require on the other hand an adjustment in the overall
  normalization.
\end{abstract}

\section{INTRODUCTION}
Ultra-peripheral collisions at the LHC provide currently the most
energetic photon-proton collisions measured so far. In particular such
reactions allow to probe the gluon distribution in the proton down to
ultra-small values of Bjorken $x$, increasing the range probed by the
HERA experiments roughly by an order of magnitude. A particular
interesting class of events are exclusive photo-production of vector
mesons $J/\Psi$ and $\Upsilon$ where the mass of the charm ($J/\Psi$)
and bottom ($\Upsilon$) quarks provide a hard mass scale, allowing
therefore for an analysis within perturbative Quantum
Chromodynamics. If current LHC data
\cite{TheALICE:2014dwa,Aaij:2013jxj,Aaij:2015kea,CMS:2016nct} are
complement with HERA data
\cite{Chekanov:2002xi,Chekanov:2004mw,Alexa:2013xxa,Aktas:2005xu,Adloff:2000vm,Breitweg:1998ki,Chekanov:2009zz}
measured at lower values of the collision energy $W$ of the effective
photon-proton collisions, these measurements allow to probe
Balitsky-Fadin-Kuraev-Lipatov (BFKL) evolution
\cite{BFKL1,Fadin:1998py} over two orders of magnitude in $W$. In
particular photo-production of $\Upsilon$ mesons provides due to the
relatively large bottom quark mass an excellent test of perturbative
low $x$ BFKL evolution. While the bottom mass is likely to completely
suppress to a very good accuracy non-linear corrections to BFKL
evolution, characteristic for the regime of saturated gluon densities,
observation of such effects should be in principle possible in
$J/\Psi$ production, where the saturation scale can reach in principle
values of the order of the charm mass, see {\it e.g.}  Reference
\cite{Armesto:2014sma}. For such studies the BFKL description provides
therefore at the very least an important benchmark in the sense that
it tests the possibility to describe the given data set without the
inclusion of non-linear high density effects. In the following we
present a few aspects of a description of both $J/\Psi$ and $\Upsilon$
photo-production data which is based on the (collinear improved) NLO
BFKL fit to HERA data of Reference
\cite{Hentschinski:2012kr,Hentschinski:2013id}. For the full details
of this study we refer the interested reader to Reference
\cite{Bautista:2016xnp}.

\section{Results \& Outlook}
\begin{figure}[p!]
 
   \parbox{.85\textwidth}{%
     \includegraphics[width=.85\textwidth]{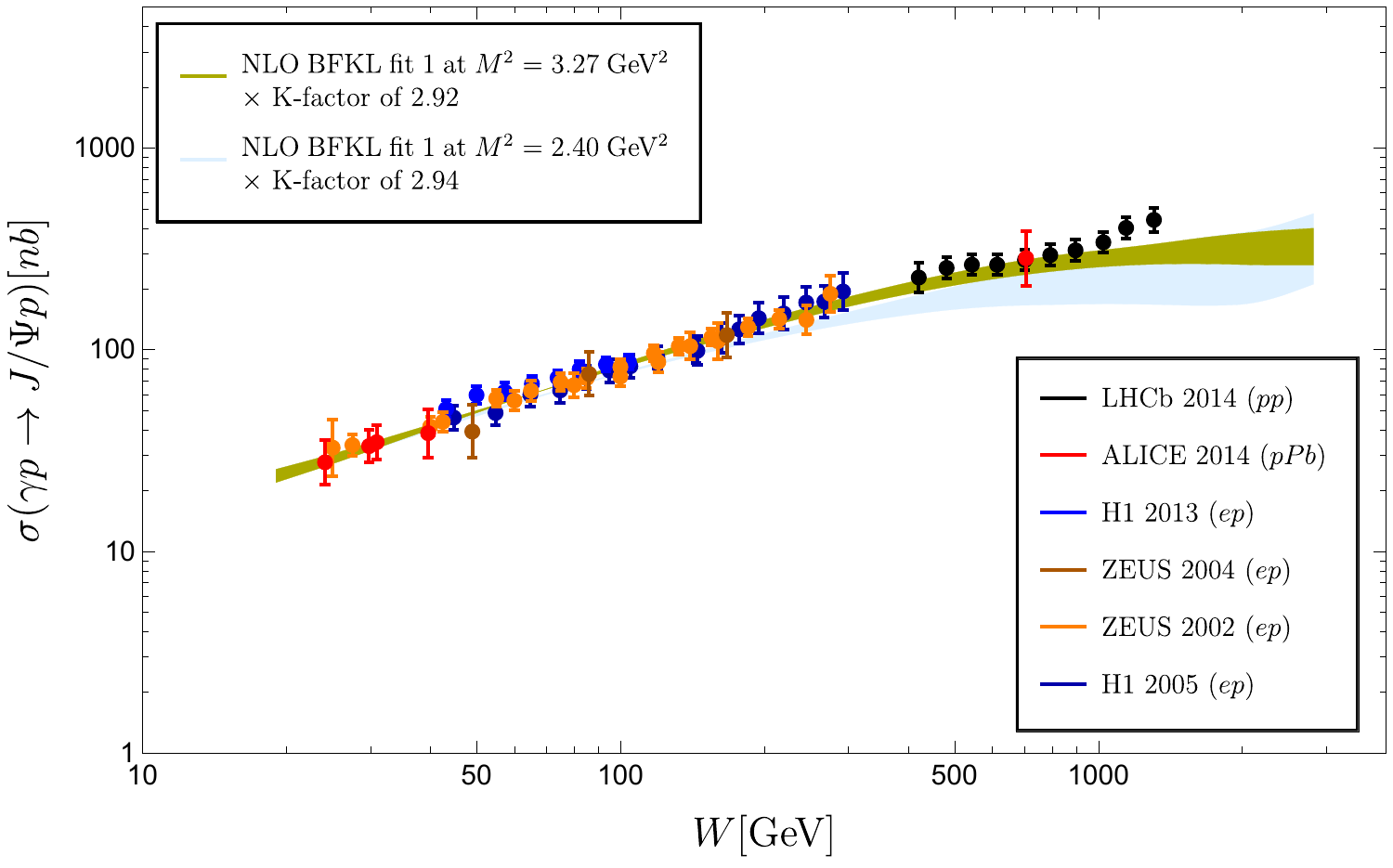}}   

   \caption{\it Energy dependence of the $J/\Psi$ photo-production
     cross-section as provided by the BFKL fit 1. The uncertainty band
     reflects a variation of the scale associated with the running
     coupling.  We also show photo-production data measured at HERA by
     ZEUS \cite{Chekanov:2002xi,Chekanov:2004mw} and H1
     \cite{Alexa:2013xxa,Aktas:2005xu} as well as LHC data obtained
     from ALICE \cite{TheALICE:2014dwa} and LHCb ($W^+$ solutions)
     \cite{Aaij:2013jxj}.}
  \label{fig:resultsJPsi}
\end{figure}

\begin{figure}[p!]
   \parbox{.85\textwidth}{%
     \includegraphics[width=.85\textwidth]{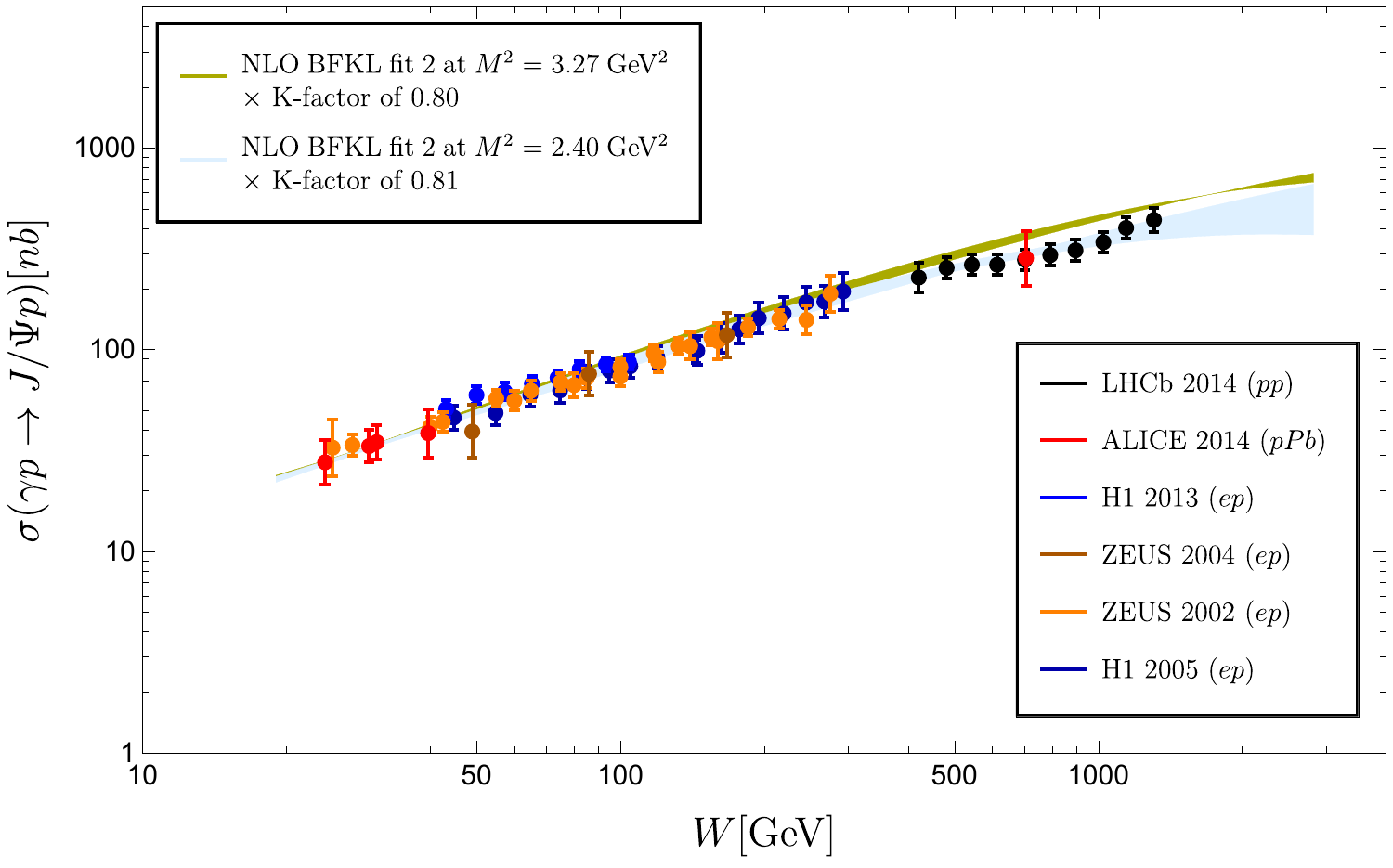}}   \\ 

  \caption{\it Energy dependence of the $J/\Psi$ photo-production
    cross-section as provided by the BFKL fit 2. For further details see Figure~\ref{fig:resultsJPsi}}
  \label{fig:resultsJPsi2}
\end{figure}

\begin{figure}[p!]
  \centering
  \includegraphics[width=.85\textwidth]{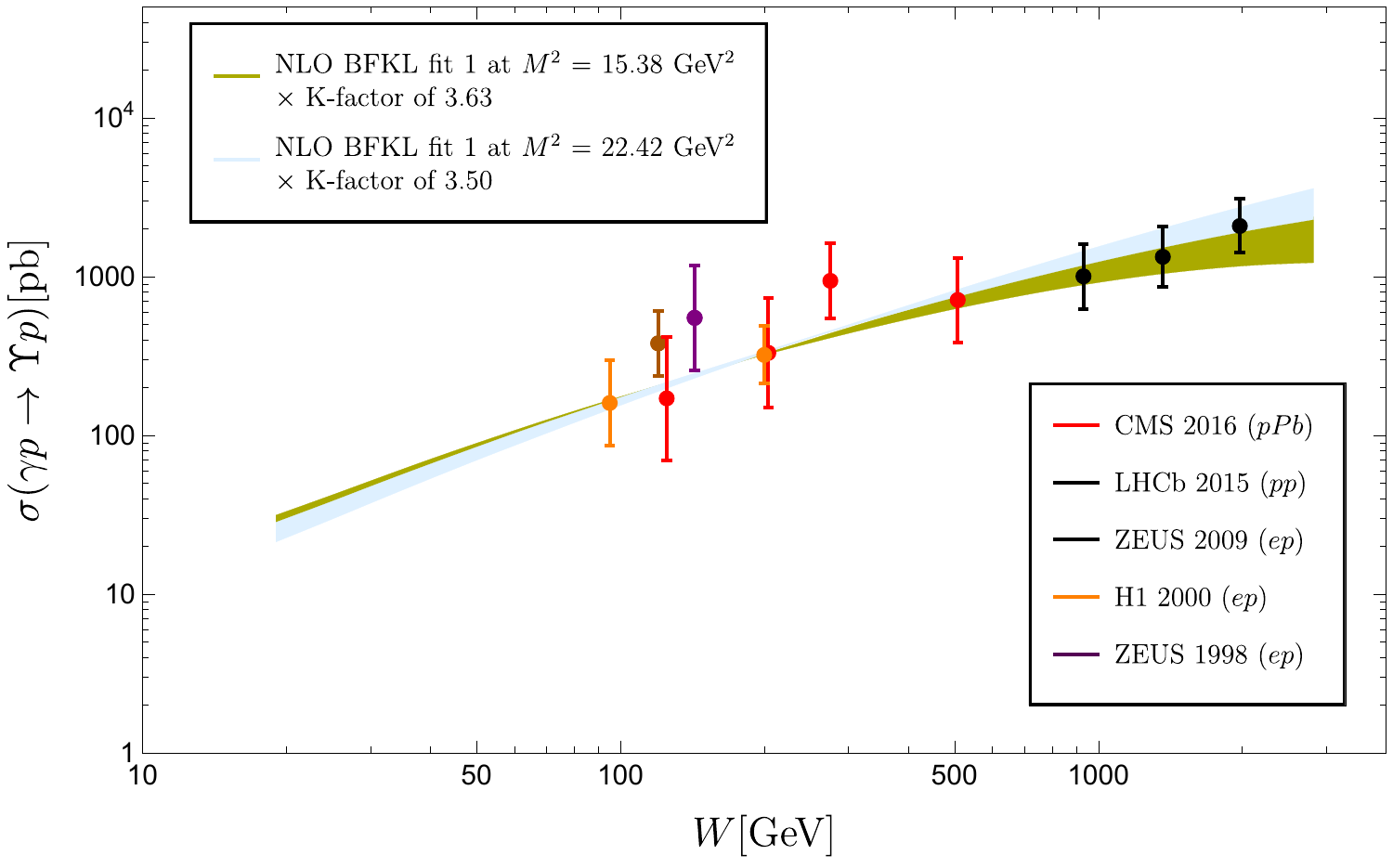}
  \caption{\it Energy dependence of the $\Upsilon$ photo-production
    cross-section as provided by the BFKL fit 1 .  The uncertainty
    band reflects a variation of the scale of the running couling.  We
    also show HERA data measured by H1 \cite{Adloff:2000vm} and ZEUS
    \cite{Breitweg:1998ki,Chekanov:2009zz} and LHC data by LHCb
    \cite{Aaij:2015kea} and CMS \cite{CMS:2016nct}.}
  \label{fig:resultsUpsilon}
\end{figure}

\begin{figure}[p!]
  \centering
   \includegraphics[width=.85\textwidth]{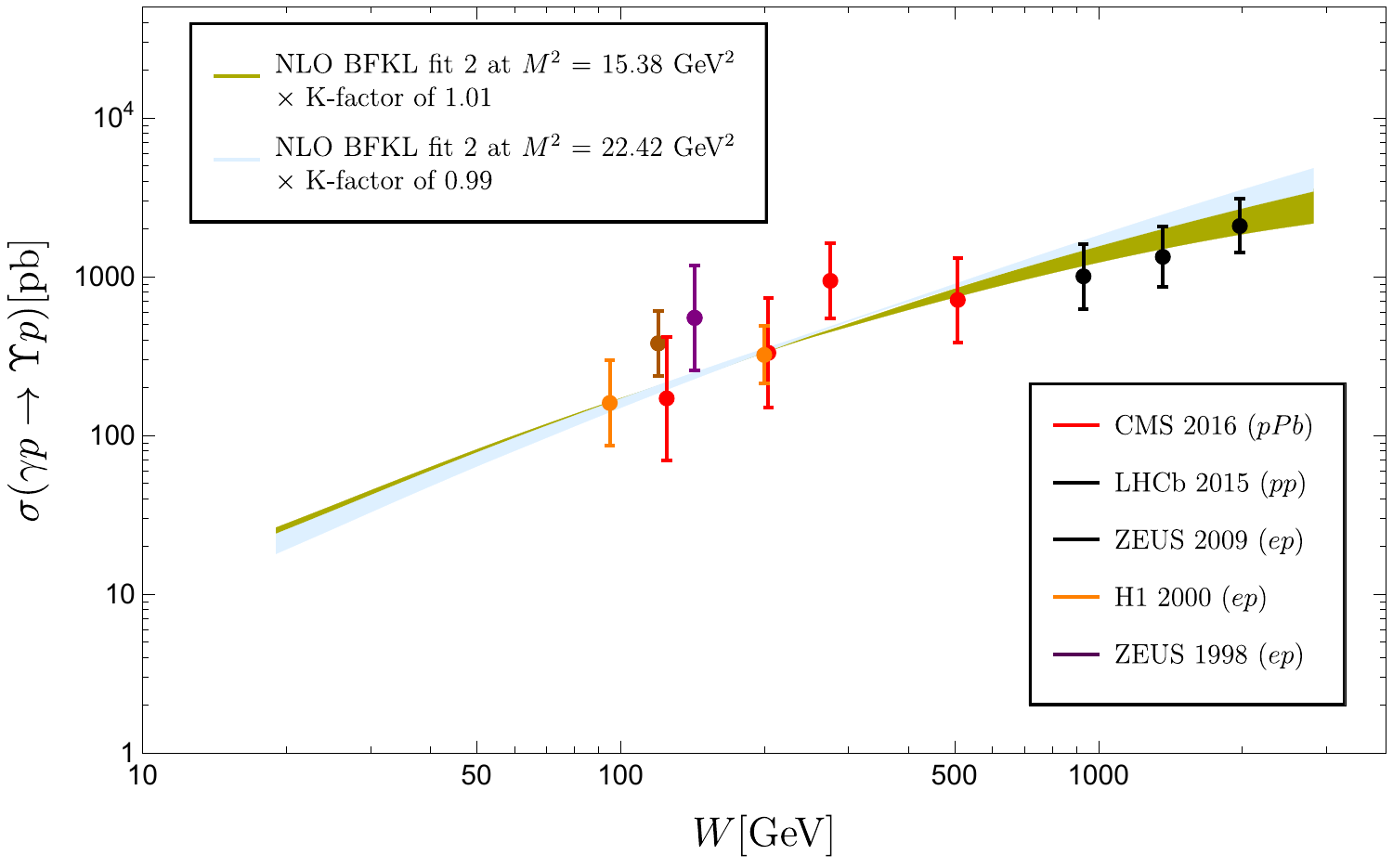}
  \caption{\it Energy dependence of the $\Upsilon$ photo-production
    cross-section as provided by the BFKL fit 2. For further details see Figure~\ref{fig:resultsUpsilon}}
  \label{fig:resultsUpsilon2}
\end{figure}

The central results of the study are shown in
Figures~\ref{fig:resultsJPsi}, \ref{fig:resultsJPsi2} ($J/\Psi$) and
Figures~\ref{fig:resultsUpsilon}, \ref{fig:resultsUpsilon2} ($\Upsilon$). While the BFKL fits provide a
very good description of the energy dependence, BFKL fit 1 requires a
relatively large adjustment in the overall normalization (of order
$3-3.5$); for fit 2 the necessary adjustment is of order one. To
improve and stabilize the BFKL description further it will be
necessary to provide a re-fit of HERA data which takes into account
heavy quark masses and possibly now available next-to-leading order
corrections to the virtual photon impact factor with massless
quarks. In addition it might be possible to achieve a more precise
description of light seaquarks in the fit, using techniques developed
in
Reference~\cite{Hautmann:2012sh,Gituliar:2015agu,Hentschinski:2016wya}. For
the $\gamma p \to V p$ cross-section this would then require the
determination of corresponding NLO corrections using for instance the
techniques developed and used in NLO calculations within high energy
factorization \cite{Hentschinski:2011tz, Hentschinski:2014lma,
  Ayala:2016lhd}.



\section{ACKNOWLEDGMENTS}
I would like to thank the organizers of DIFFRACTION 2016 for a wonderful workshop. I further acknowledge support by  CONACyT-Mexico grant number CB-2014-241408.


\nocite{*}
\bibliographystyle{aipnum-cp}%

\end{document}